\documentclass[reprint, amsmath, amssymb, aps, prp, superscriptaddress]{revtex4-2}
\usepackage{times}
\usepackage{helvet}
\usepackage{courier}
\usepackage{graphicx}
\usepackage{dcolumn}
\usepackage{bm}
\usepackage[colorlinks=true, linkcolor=blue, citecolor=blue, urlcolor=blue]{hyperref}

\usepackage{caption}
\usepackage{subcaption}
\usepackage{color}
\usepackage{float}
\usepackage{booktabs}

\begin{document}

\title{Parallel iQCC Enables 200 Qubit Scale Quantum Chemistry on Accelerated Computing Platforms Surpassing Classical Benchmarks in Ruthenium Catalysts}

\author{Seyyed Mehdi Hosseini Jenab}
\affiliation{OTI Lumionics Inc., 3415 American Dr., Mississauga, ON L4V 1T4, Canada}

\author{Brandon Henderson}
\affiliation{OTI Lumionics Inc., 3415 American Dr., Mississauga, ON L4V 1T4, Canada}

\author{Scott N. Genin}
\email{scott.genin@otilumionics.com}
\thanks{Corresponding author}
\affiliation{OTI Lumionics Inc., 3415 American Dr., Mississauga, ON L4V 1T4, Canada}

\date{\today}

\begin{abstract}
 We introduce a parallel, GPU-accelerated implementation of the iterative qubit coupled cluster (iQCC) method that overcomes the exponential growth of the transformed Hamiltonian---the principal bottleneck for classical emulation of quantum chemistry circuits. By distributing Hamiltonian terms across compute nodes via bit-wise partitioning and offloading Pauli contractions to GPUs, we achieve speedups exceeding two orders of magnitude over the serial CPU approach. Crucially, iQCC confines the variational evolution to a classically simulable operator subspace by selecting entanglers exclusively from the Direct Interaction Space, which guarantees non-vanishing energy gradients at every iteration and thereby naturally avoids the barren-plateau phenomenon that renders highly expressive quantum circuits untrainable. Leveraging these algorithmic and hardware advances, we simulate electronic-structure Hamiltonians for industrially relevant ruthenium catalysts in the 100--124 qubit regime, completing full ground-state calculations on NVIDIA GPUs in the ranges of 1.2 - 45  hrs and surpassing the accuracy of Density Matrix Renormalization Group. These results effectively de-quantize a significant portion of the NISQ roadmap: quantum advantage for chemistry is often assumed to emerge beyond ${\sim}50$ qubits, yet our work demonstrates that this frontier lies significantly further---potentially past 200 qubits---reshaping expectations for where genuine quantum advantage may first appear.
\end{abstract}

\maketitle


\section{Introduction}
Quantum simulation of molecular electronic structure is widely regarded as a flagship application of quantum computing. Computing the ground-state energy of molecules from the Schr\"{o}dinger equation under the Born--Oppenheimer approximation scales exponentially with system size for exact classical methods, making it a natural candidate for quantum speedup. Since the first demonstration of quantum computational supremacy in 2019 \cite{google2019_supremacy}, significant advances have been made in quantum error correction \cite{google2024_errorcorr,quantinuum2025}, while classical emulation frameworks have simultaneously pushed the boundaries of what can be simulated without quantum hardware \cite{aws2023_sparse,google2023_tensor,npj2023_nasc}. A central question thus persists: at what system size will quantum devices deliver practical advantage over classical computation for chemistry? The threshold is conventionally placed near 50 qubits, a boundary derived from the memory limits of state-vector simulation and often cited as the onset of genuine quantum advantage \cite{ibm2025_sciadv_qcsc}. However, this boundary is increasingly contested by advances on both the classical and quantum-classical fronts.

On the classical side, several recent developments have pushed emulation capabilities well beyond the 50-qubit mark. Truncated sparse tensor simulation (TruSTS) has extended approximate state-vector methods to $N{=}64$ qubits with near-constant runtime scaling \cite{miller2026_trusts}, though this still falls short of the system sizes routinely encountered in catalytic and materials chemistry. Tensor-network methods have made parallel strides: mixed-precision DMRG on NVIDIA Blackwell GPUs achieves chemical accuracy for active spaces as large as CAS(113,76) \cite{brower2025_blackwell_dmrg}, and GPU-accelerated 2D tensor networks can simulate quantum circuits well beyond 50 qubits \cite{rudolph2025_2dtn}. However, these approaches remain fundamentally governed by bond-dimension scaling, which grows steeply for strongly correlated transition-metal systems. On the quantum-classical front, the most ambitious calculation to date coupled an IBM Heron processor with the Fugaku supercomputer to tackle electronic-structure problems up to 77 qubits, arguing that chemistry near the 100-qubit mark requires tightly integrated quantum-classical workflows \cite{ibm2025_sciadv_qcsc}. These results collectively suggest that classical tractability extends further than previously assumed, but the precise boundary---particularly for industrially relevant transition-metal chemistry---remains an open question.

The choice of quantum algorithm is central to this question. Early hopes were pinned on quantum phase estimation (QPE) \cite{abrams1999_qpe_prl}, which promises asymptotic efficiency but demands deep circuits unsuited to noisy intermediate-scale quantum (NISQ) hardware. The variational quantum eigensolver (VQE) offered a more practical alternative, demonstrating proof-of-concept ground-state calculations for simple molecules \cite{peruzzo2014_vqe}. However, even with extensive Hamiltonian simplification and ansatz optimization, current hardware noise levels prevent VQE from producing chemically reliable energies for molecules as modest as benzene \cite{carreras2025_vqe_limitations}. Standard VQE approaches further suffer from the barren-plateau phenomenon: as circuits become increasingly expressive, the optimization landscape flattens exponentially, rendering gradient-based training intractable \cite{Larocca_2025}. These limitations motivated qubit-space formulations that address trainability by construction. The qubit coupled cluster (QCC) method constructs trial states directly in the qubit basis and restricts the ansatz to generators from the Direct Interaction Space (DIS)---a subset of Pauli operators guaranteed to produce non-zero energy gradients \cite{ryabinkin2018_qcc}. Its iterative extension (iQCC) applies a sequence of canonical transformations, achieving chemically accurate results with shallow circuits while maintaining well-conditioned optimization landscapes at every iteration \cite{ryabinkin2019_iqcc}.

Despite these algorithmic advances, scaling remains the critical bottleneck. Each iQCC iteration generates an exponential growth in the number of Pauli terms in the effective Hamiltonian, driving up classical memory and runtime requirements. Several strategies have been proposed to mitigate this burden, including a posteriori perturbative corrections \cite{ryabinkin2020_a_posteriori}, linear-combination refinements \cite{ryabinkin2023_ilcap}, polynomial optimization schemes \cite{ryabinkin2025_optimization}, and sparse simulation methods \cite{steiger2024_sparse,miller2026_trusts}. Nevertheless, most implementations remain serial, reinforcing the widespread assumption that classical emulation of quantum chemistry is intractable beyond $\sim$50~qubits.

In this work, we directly challenge that assumption. By distributing Hamiltonian terms across compute nodes via bit-wise partitioning and offloading Pauli contractions to GPUs, we develop a parallel, GPU-accelerated iQCC framework that delivers speedups exceeding two orders of magnitude over serial implementations. This approach overcomes the exponential operator growth, completing variationally bounded simulations of electronic-structure Hamiltonians in the 100--124 qubit regime within hours on commodity GPU hardware. Crucially, by confining the variational evolution to a subspace that naturally avoids barren plateaus, the iQCC optimization remains classically tractable. These results demonstrate that, at least for electronic structure, the frontier of classical tractability extends well beyond the conventionally assumed 50-qubit boundary---suggesting that genuine quantum advantage may not emerge until system sizes approach 200 qubits or more.

\section{Methods}
\begin{figure*}
    \centering
    \includegraphics[width=1.0\linewidth]{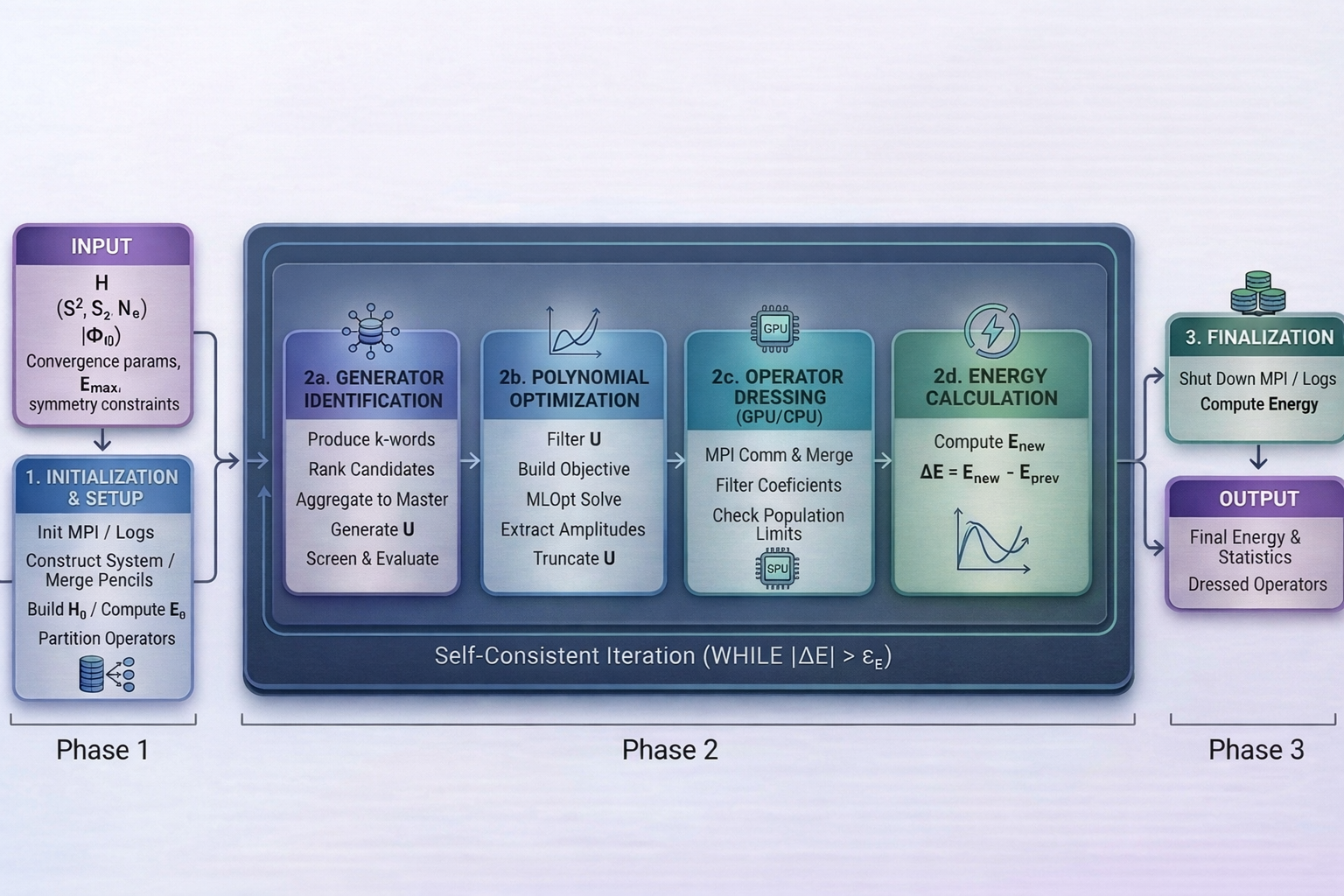}
    \caption{Overview of the parallel iQCC optimization workflow. After initialization and MPI distribution of the Hamiltonian (Step~1), the algorithm enters an iterative loop comprising: (2a)~generator identification from the Direct Interaction Space via gradient screening, (2b)~optional polynomial optimization of entangler amplitudes on GPUs, (2c)~GPU-accelerated operator dressing of the Hamiltonian, and (2d)~convergence checking against energy and iteration thresholds. The loop terminates when convergence criteria are met, followed by finalization (Step~3). Color coding distinguishes input/output (blue), initialization and finalization (red), the iterative core (orange/green), and GPU-accelerated stages (purple).}
    \label{fig:workflow}
\end{figure*}
\subsection{Theoretical Framework of the Iterative Qubit Coupled Cluster (iQCC) Method}

The QCC method begins with the second-quantized electronic Hamiltonian, defined as:
\begin{equation}
\hat{H}_{e} = \sum_{pq} h_{pq} \hat{a}_{p}^{\dagger} \hat{a}_{q} + \frac{1}{2} \sum_{pqrs} g_{pqrs} \hat{a}_{p}^{\dagger} \hat{a}_{q}^{\dagger} \hat{a}_{r} \hat{a}_{s}
\end{equation}
where $\hat{a}_{p}^{\dagger}$ and $\hat{a}_{q}$ are the fermionic creation and annihilation operators, and $h_{pq}$ and $g_{pqrs}$ are the one- and two-electron integrals. Through a fermion-to-qubit mapping, such as the Jordan-Wigner or Bravyi-Kitaev transformation, this Hamiltonian is expressed as a linear combination of Pauli strings:
\begin{equation}
\hat{H} = \sum_{k} C_{k} \hat{P}_{k}
\end{equation}
where $C_{k}$ are real coefficients and $\hat{P}_{k} \in \{\hat{I}, \hat{X}, \hat{Y}, \hat{Z}\}^{\otimes N_q}$ are multi-qubit Pauli words.

The reference state is a qubit mean-field (QMF) wavefunction, a tensor product of single-qubit coherent states:
\begin{equation}
|\Omega\rangle = \bigotimes_{j=1}^{N_q} \left( \cos\frac{\theta_{j}}{2} |0\rangle_{j} + e^{i\phi_{j}} \sin\frac{\theta_{j}}{2} |1\rangle_{j} \right)
\end{equation}
where $\theta_{j}$ and $\phi_{j}$ are variational Bloch angles for the $j$-th qubit. Electron correlation is systematically introduced by applying a unitary operator directly in the qubit space, forming the QCC ansatz:
\begin{equation}
|\Psi(\tau, \Omega)\rangle = \hat{U}(\tau) |\Omega\rangle = \prod_{k=1}^{N_{ent}} \exp\left(-i\frac{\tau_{k}}{2} \hat{P}_{k}\right) |\Omega\rangle
\end{equation}
where $\tau_{k}$ are variational amplitudes and $\hat{P}_{k}$ are multi-qubit Pauli operators serving as entanglement generators.

A significant mathematical advantage of the QCC formulation over traditional fermionic unitary coupled-cluster methods is that the involutory property of Pauli strings ($\hat{P}_k^2 = \hat{I}$) truncates the Baker-Campbell-Hausdorff expansion. The similarity-transformed (dressed) Hamiltonian for a single generator can thus be analytically evaluated in closed form:
\begin{equation}
\begin{split}
\hat{H}_{d}(\tau_k) &= e^{i\frac{\tau_k}{2}\hat{P}_k} \hat{H} e^{-i\frac{\tau_k}{2}\hat{P}_k} \\
  &= \hat{H} - \frac{i}{2} \sin(\tau_k) [\hat{H}, \hat{P}_k] \\
  &\quad + \frac{1}{2} (1 - \cos(\tau_k)) \hat{P}_k [\hat{H}, \hat{P}_k]
\end{split}
\end{equation}

In the iterative extension (iQCC), the Hamiltonian is sequentially dressed to incorporate correlation effects while maintaining a shallow quantum circuit at each step:
\begin{equation}
\hat{H}^{(n+1)} = \hat{U}^{\dagger}(\tau_{opt}^{(n)}) \hat{H}^{(n)} \hat{U}(\tau_{opt}^{(n)})
\end{equation}

To navigate the exponentially large generator space efficiently, iQCC utilizes a gradient-based screening protocol. The zero-amplitude energy gradient with respect to a candidate generator $\hat{P}_{\alpha}$ provides the direction of steepest energy decrease:
\begin{equation}
g_{\alpha} = \left. \frac{\partial E}{\partial \tau_{\alpha}} \right|_{\tau_{\alpha}=0} = -\frac{i}{2} \langle \Omega | [\hat{H}, \hat{P}_{\alpha}] | \Omega \rangle = \text{Im} \langle \Omega | \hat{H} \hat{P}_{\alpha} | \Omega \rangle
\end{equation}
Operators yielding non-zero gradients constitute the Direct Interaction Space (DIS). Ranking these gradients allows the algorithm to systematically select the most energetically impactful entanglers for subsequent amplitude optimization and Hamiltonian dressing.

\subsection{Parallelization Strategy}

To overcome the memory and runtime bottlenecks of Hamiltonian growth, we developed a parallel iQCC framework. Because inter-GPU communication is the dominant bottleneck in multi-GPU quantum simulation, with interconnect advances yielding larger speedups than improvements in GPU compute alone \cite{brown2025_multigpu_network}, the central design principle is that Hamiltonian terms are \emph{never gathered} on a single node: each compute core is responsible for a disjoint subset of terms, and all operations are performed locally whenever possible. We employ a bit-wise partitioning strategy that distributes terms based on specific bits in their binary representation, ensuring that newly generated terms from the dressing procedure can be deterministically routed to their correct partitions, with communication required only in restricted, pairwise cases. This design minimizes all-to-all communication and allows the algorithm to scale efficiently with the number of processors. To maintain high utilization as the Hamiltonian grows unevenly across partitions, we implement fine-grained partitioning and dynamic load balancing, where partitions can be reassigned to underutilized cores without disrupting correctness.

\subsection{GPU Acceleration}

Expectation-value evaluation and Hamiltonian dressing are highly parallelizable, consisting of large numbers of independent Pauli contractions. We exploit this structure by offloading these operations to GPUs. Pauli terms are encoded as binary strings of $X$ and $Z$ components, enabling efficient bitwise operations in CUDA kernels. Sparse indexing and warp-synchronous reductions maximize throughput and memory locality. This GPU implementation yields an additional order-of-magnitude performance improvement relative to CPU-only parallelization.

\subsection{Polynomial Optimization of Amplitudes}

Amplitude optimization becomes a further bottleneck as the number of entanglers grows into the thousands. We employ a polynomial optimization scheme~\cite{ryabinkin2025_optimization} that approximates the QCC unitary by a truncated symmetric-polynomial expansion. For a set of $N$ entanglers, the full expansion contains $2^N$ terms; truncation at order $K$ reduces this to $\mathcal{O}(N^K)$, providing systematic control over the accuracy-efficiency tradeoff. The resulting objective function remains smooth and differentiable, making it compatible with gradient-based optimizers. In practice, values of $K=2$--$6$ recover sub-millihartree accuracy while allowing simultaneous optimization over hundreds of thousands of entanglers. The expansion is parallelized across nodes: each core constructs its assigned portion of the polynomial expansion, evaluates local contributions to energy and gradients, and communicates only aggregated results to the optimizer.

\subsection{Complexity and Scaling}

The dominant cost in iQCC is the exponential growth of Hamiltonian terms under repeated dressing: each iteration can increase the term count by up to a factor of $3/2$ \cite{ryabinkin2019_iqcc}, so that after $n$ iterations the Hamiltonian may contain $\mathcal{O}((3/2)^n \, M_0)$ terms, where $M_0$ is the initial count. Our framework controls this growth at every level of the computation. Bit-wise partitioning distributes the $M$ terms across $P$ processors without duplication, yielding $\mathcal{O}(M/P)$ memory per node. Communication during dressing is restricted to pairwise exchanges between nodes whose partitions differ in the flipped bits; energy and gradient evaluations require only a single $\mathcal{O}(\log P)$ global reduction. The sortless dressing algorithm (Appendix~\ref{appendix:sortless}) further reduces the per-step cost by replacing a conventional $\mathcal{O}(M \log M)$ global sort of newly generated terms with a partition-aware merge that operates in $\mathcal{O}(M)$ time. On the optimizer side, truncating the symmetric-polynomial expansion of the QCC unitary at order $K$ reduces the number of operator terms from $2^N$ to $\sum_{k=0}^{K}\binom{N}{k} = \mathcal{O}(N^K)$ for $N$ entanglers \cite{ryabinkin2025_optimization}; the associated Hessian construction is linear in the number of Hamiltonian terms, enabling the simultaneous optimization of millions of amplitudes. Meanwhile, the DIS construction guarantees that the number of candidate generators scales polynomially with system size rather than exponentially \cite{ryabinkin2019_iqcc}. In practice, these scalings allow Hamiltonians exceeding $10^9$ Pauli terms and simultaneous optimization of millions of entanglers, extending classical tractability to the 200-qubit regime on commodity GPU hardware.

\subsection{Computational Workflow}

The complete workflow of a parallel iQCC calculation proceeds as follows (see Figure~\ref{fig:workflow} for an overview):
\begin{enumerate}
    \item Construct the qubit Hamiltonian via a fermion-to-qubit transformation (e.g., Jordan-Wigner or Bravyi-Kitaev) and distribute terms across compute nodes using bit-wise partitioning.
    \item Initialize the qubit mean-field (QMF) reference state.
    \item Enter the iterative loop: identify candidate entanglers from the DIS, optimize their amplitudes with the polynomial optimizer (aggregating energy and gradient information across all nodes), and dress the Hamiltonian.
    \item Repeat step~3 until the energy change falls below a specified tolerance (e.g., $10^{-6}$~Hartree) or a maximum iteration count is reached.
\end{enumerate}

\subsection{Hardware Specifications}
All calculations were performed on one of two hardware configurations:
\begin{itemize}
  \item \textbf{V100 node:} 4$\times$ NVIDIA V100 GPUs (32~GB HBM2 each), 2$\times$ Intel Xeon Gold 6248 processors (40 cores total, 2.50~GHz), NVIDIA driver 535.183.01, CUDA 12.2.
  \item \textbf{B200/B300 nodes:} NVIDIA B200 or B300 GPUs provisioned via the NVIDIA Brev cloud platform, paired with AMD EPYC 9575F processors (64 cores, 2.0~GHz), NVIDIA driver 580.126.09.
\end{itemize}
Throughout this work, the term ``CPU'' denotes individual physical processing cores rather than discrete multi-core processor sockets.


\section{Results}
\subsection{Hydrogen molecule benchmark}

To validate the accuracy and efficiency of the proposed polynomial optimization scheme,
we first consider the hydrogen molecule (H$_2$) across several basis sets
(STO-3G, cc-pVDZ, cc-pVTZ, and aug-cc-pVTZ).
This system is traditionally used as a benchmark for quantum chemistry algorithms
because its small size allows exact reference calculations while still providing a
non-trivial growth in the number of qubits and entangling operations with increasing basis size\cite{mccaskey2019quantum,mihalikova2022best}.

In Table~\ref{tab:H2}, we report both the number of spatial orbitals ($N_{\text{basis}}$)
and the corresponding qubit counts.
Each spatial orbital supports both $\alpha$ and $\beta$ spin projections, giving rise to two spin orbitals.
Because fermion-to-qubit transformations (Jordan--Wigner, Bravyi--Kitaev, etc.)
map each spin orbital to one qubit, the number of qubits used in iQCC is
$N_{\text{qubits}} = 2 \times N_{\text{basis}}$.
For H$_2$, this corresponds to 4, 20, 56, and 92 qubits in the four basis sets considered.
While these are still modest numbers, they already illustrate how circuit complexity
and entangler counts grow rapidly with basis size.

For these H$_2$ benchmarks, no iterative dressing of the Hamiltonian was employed. Instead, the input qubit Hamiltonian was taken as is, and polynomial optimization was performed directly. In this simple system, the polynomial expansion alone suffices to reproduce the full configuration interaction (FCI) energy to within $10^{-5}$~Ha, which exceeds the standard chemical accuracy threshold ($1.6$ mHa) of a given basis set. This demonstrates the intrinsic efficiency of the polynomial scheme: even without dressing, it achieves numerically exact FCI agreement at negligible computational cost. For larger, strongly correlated molecules, dressing iterations are necessary to redistribute Hamiltonian weight among entanglers, but the present results show that the polynomial method captures essential physics in the simplest nontrivial cases.

All iQCC+poly calculations reproduce the FCI energy to within $10^{-5}$~Ha, well beyond the threshold of chemical accuracy. For the H$_2$ molecule, the QCC Ansatz will only generate entanglers acting upon only 2 or 4 qubits since this corresponds to either single excitations and double excitations since the molecule contains only 2 electons in the active space. The computational timings are several orders of magnitude smaller than those reported by Shang \emph{et al.}~\cite{shang2023massive}, who employed an MPS-based VQE emulator on the Sunway supercomputer. Their simulations required between $0.12$~s and $1564.5$~s per iteration, corresponding to wall times ranging from hours to weeks, whereas our calculations complete in milliseconds. All results reported here were obtained on a workstation equipped with dual Intel\textsuperscript{\textregistered} Xeon\textsuperscript{\textregistered} E5-2650 v3 processors (40 physical cores, 80 threads, 2.30~GHz base frequency), using 16 cores for the calculations. Additional tests on two NVIDIA GPUs yielded nearly identical timings due to the small problem size.


The dramatic difference in performance highlights the central advantage of the
polynomial optimization scheme: by analytically restructuring the QCC amplitudes
into polynomial expansions, we avoid the tensor-network overhead inherent in
MPS-based emulators. Although GPU-accelerated tensor-network methods continue to
improve---most recently through mixed-precision DMRG on Blackwell hardware
\cite{brower2025_blackwell_dmrg}---they remain fundamentally constrained by
bond-dimension scaling, a limitation absent from the polynomial optimization scheme.
This enables speedups of $10^{4}$--$10^{7}$ while retaining exact agreement with
reference FCI energies.

\begin{table*}[tbp]
\centering
\caption{Comparison of H$_2$ (R$=0.75A$) ground-state energies and timings obtained with
iQCC+poly optimization (this work) and MPS--VQE emulation on the Sunway supercomputer~\cite{shang2023massive}.
CNOT counts are estimated as $2(n-1)$ per $n$-qubit entangler.}
\label{tab:H2}
\footnotesize
\setlength{\tabcolsep}{6pt}
\renewcommand{\arraystretch}{1.15}
\begin{tabular}{lcccc}
\hline\hline
 & STO-3G & cc-pVDZ & cc-pVTZ & aug-cc-pVTZ \\
\hline
$N_{\text{basis}}$ (spatial orbitals) & 2 & 10 & 28 & 46 \\
$N_{\text{qubits}}$ (spin orbitals)   & 4 & 20 & 56 & 92 \\
FCI Energy (Ha)             & $-1.13712$  & $-1.16359$   & $-1.17230$   & $-1.172604$ \\
QCC/iQCC Energy (Ha)        & $-1.13712$  & $-1.16359$   & $-1.17230$   & $-1.172604$ \\
Entanglers (total)          & $1$         & $21$         & $111$        & $317$ \\
\quad 4-qubit               & $1$         & $17$         & $101$        & $297$ \\
\quad 2-qubit               & $0$         & $4$          & $10$         & $20$ \\
CNOT gates                  & $6$         & $110$        & $626$        & $1822$ \\
Timing (iQCC+poly, 16 CPU) (s) & $1.47{\times}10^{-4}$ & $1.10{\times}10^{-3}$ & $4.25{\times}10^{-3}$ & $1.72{\times}10^{-2}$ \\
Timing (Sunway MPS--VQE) (s)   & $2.16$      & $1.112{\times}10^{3}$ & $8.75{\times}10^{4}$ & $1.059{\times}10^{6}$ \\
Speedup factor                 & $1.47{\times}10^{4}$ & $1.01{\times}10^{6}$ & $2.06{\times}10^{7}$ & $6.17{\times}10^{7}$ \\
\hline\hline
\end{tabular}
\end{table*}

\subsection{Ruthenium Catalyst Carbon Sequestration Cycle Benchmark Systems}

Eight ruthenium-based complexes from the CO$_2$ fixation catalytic cycle (I, II, II-III, V, VIII, VIII–IX, IX, and XVIII) \cite{GHGcatalyst2015} were selected to evaluate the accuracy and performance of the \texttt{iQCC} implementation on both NVIDIA V100 GPUs and a subset of the systems were implemented on NVIDIA B200 GPUs. The 1 and 2 electron integral files were obtained from von Burg et al. \cite{von2021quantum} and as noted, the structures were first optimized using Density Functional Theory (DFT) with M06-L/def2-SVP. The initial Hartree--Fock (HF) orbitals were prepared by using the ANO-RCC-VTZP basis set on the non-metal atoms and ANO-RCC-VQZP on the Ruthenium atom which where then followed by using CASSCF to generate the final set of electron integrals with a 10$^{-4}$ Cholesky Decomposition threshold. These ruthenium complexes exemplify the ``class-2'' multi-configurational electronic structures recently identified as the hardest targets for classical methods and the natural benchmarks for quantum algorithms \cite{morchen2024_classification}. The target reference energies were taken from reported DMRG-CI calculations \cite{von2021quantum}. A Qubit mean-field calculation \cite{ryabinkin-qmf-2018} was performed with the 1 and 2 electron integrals that were derived from the HF canonical molecular orbitals (system II) to confirm that the Qubit mean-field energies agreed with the reported HF energies.

\subsubsection{Accuracy of the iQCC Energies}

\begin{table*}[ht]
\centering
\caption{Summary of iQCC results on V100 GPUs for complexes I, II, II-III, and V. Energies are in Ha, deviations in~mHa. DMRG-CI values are taken from von Burg \textit{et al.} \cite{von2021quantum}}
\label{tab:v100-results1}
\begin{tabular}{lcccc}
\hline
 & I & II & II-III & V \\ \hline
CAS & (48e,52o) & (70e,62o) & (74e,65o) & (68e,60o) \\
Total iQCC Steps & 100 & 200 & 400 & 231 \\
Deep Dive Step & 100 & 200 & 350 & 230 \\
Maximum Terms (10$^6$) & 100.0 & 165.0 & 500.0 & 165.00 \\
Number of Entanglers (10$^6$) & 2.0 & 2.5 & 2.0 & 2.5 \\
DMRG-CI (Ha)\cite{von2021quantum}  & -7361.461381 & -7318.123548 & -7318.099062 & -7548.296446 \\
iQCC Variational (Ha) & -7361.4559985 & -7318.125513 & -7318.099392 & -7548.297164 \\
Difference from DMRG (mHa) & \textbf{5.382} & \textbf{-1.965} & \textbf{-0.330} & \textbf{-0.718} \\
PT/EN2 (Ha) & -7361.459952 & -7318.129564 & -7318.105099 & -7548.29949 \\
Difference from DMRG (mHa) & 1.429 & -6.016 & -6.037 & -3.045 \\
$\langle S_z \rangle$ (10$^{-5}$) & 4.412 & 3.095 & 1.185 & 5.451 \\
Optimization Time (min) & 29.572 & 127.90 & 218.46 & 109.31 \\
Total Time (hrs) & 0.947 & 3.799 & 18.343 & 3.655 \\
\hline
\end{tabular}
\end{table*}

\begin{table*}[ht]
\centering
\caption{Summary of iQCC results on V100 GPUs for complexes VIII, VIII-IX, IX, XVIII. Energies are in Ha, deviations in~mHa. DMRG-CI values are taken from von Burg \textit{et al.} \cite{von2021quantum}}
\label{tab:v100-results2}
\begin{tabular}{lcccc}
\hline
 & VIII & VIII-IX & IX & XVIII \\ \hline
CAS & (76e,65o) & (72e,59o) & (68e,62o) & (64e,56o) \\
Total iQCC Steps & 200 & 175 & 600 & 200 \\
Deep Dive Step & 200 & 150 & 600 & 150 \\
Maximum Terms (10$^6$) & 100.0 & 165.0 & 1000.0 & 165.00 \\
Number of Entanglers (10$^6$) & 0.75 & 3.0 & 2.0 & 2.5 \\
DMRG-CI (Ha)\cite{von2021quantum} & -7319.23473 & -7318.06620 & -7318.30305 & -7475.43923 \\
iQCC Variational (Ha) & -7319.23650 & -7318.303087 & -7318.30362 & -7475.44068 \\
Difference from DMRG (mHa) & \textbf{-1.766} & \textbf{-1.019} & \textbf{-0.042} & \textbf{-1.451} \\
PT/EN2 (Ha) & -7319.24463 & -7318.07211 & -7318.31021 & -7475.44269 \\
Difference from DMRG (mHa) & -9.901 & -5.913 & -7.703 & -3.461 \\
$\langle S_z \rangle$ (10$^{-5}$) & -46.96 & 11.92 & -7.95 & 4.56 \\
Optimization Time (min) & 17.687 & 171.063 & 220.205 & 92.28 \\
Total Time (hrs) & 2.341 & 4.345 & 45.141 & 3.140 \\
\hline
\end{tabular}
\end{table*}

\begin{figure}[ht]
\centering
 \includegraphics[width=\linewidth]{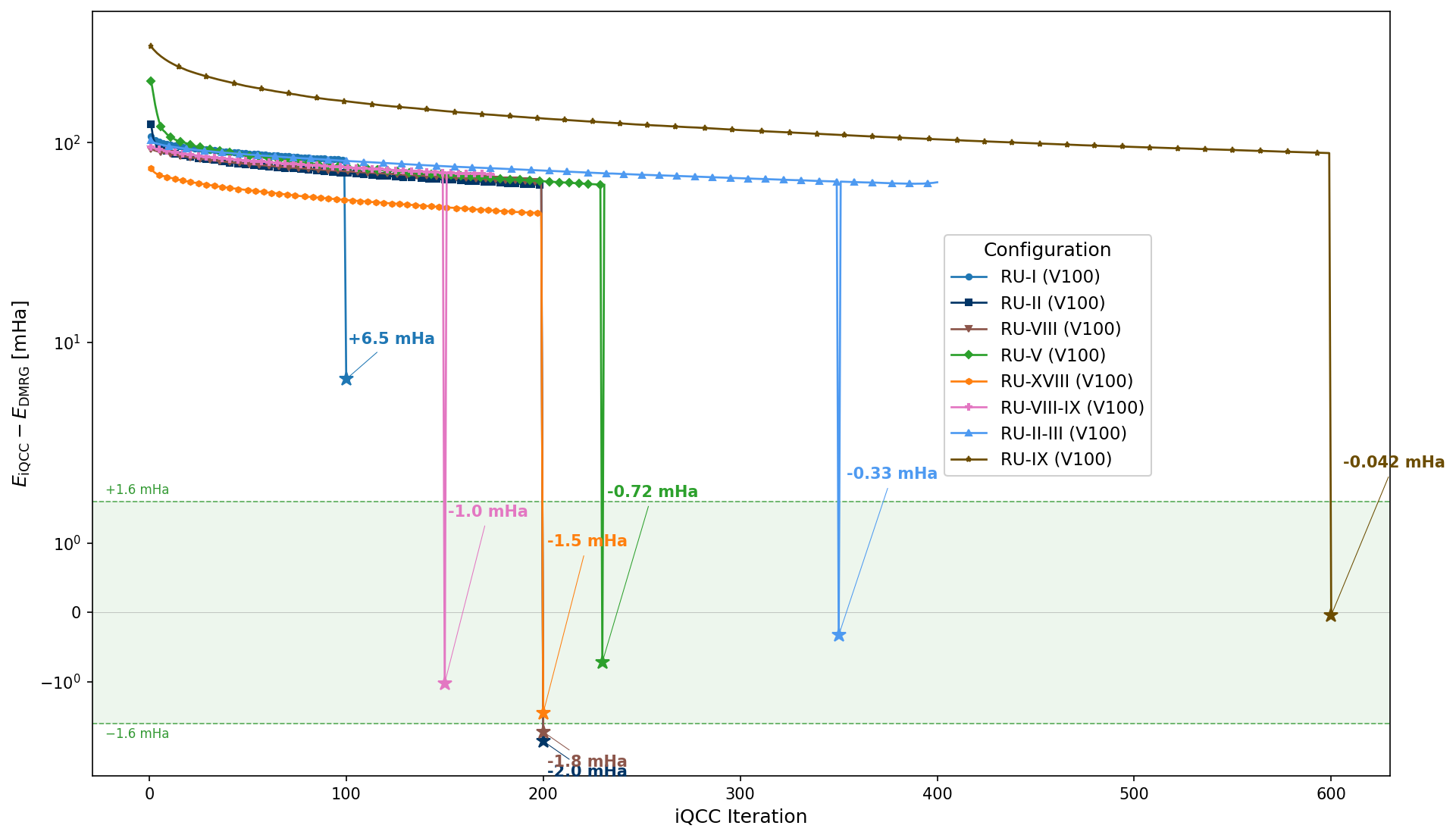}
 \caption{Energy convergence for the eight Ruthenium catalyst systems optimized using iQCC on 4 NVIDIA V100s}
 \label{fig:energy_ru}
\end{figure}

Across all systems, the polynomially optimized iQCC energies are lower than the DMRG-CI energy except for system I (Figure \ref{fig:energy_ru}). The perturbative PT/EN2 correction leads to slightly lower energies (by 2–6~mHa) and is within 1.6 mHa of the DMRG-CI energy for system I. The small deviations of $\langle S_z \rangle$ ($<2\times10^{-4}$) and $\langle N \rangle$ ($<3\times10^{-3}$) confirm proper spin and particle-number conservation during the optimization (Tables \ref{tab:v100-results1} and \ref{tab:v100-results2}). In general the PT energies are more sensitive to the coefficient cutoff and the term limit, but the variational value does not appear so. A main benefit we have seen here, but not in previous manuscripts is that when the iQCC has hit a 'saturation point' the energy starts to increase with respect to increased dressings (Figure \ref{fig:energy_ru}) which indicates the Hamiltonian has deviated to far from the isospectral transformation that iQCC intends \cite{ryabinkin2018_qcc}. This is a useful insight for a practitioner as it gives a quantifiable insight into when to being the final 'deep-dive' optimization where millions of entanglers are optimized.

The results show that iQCC can exceed DMRG-CI energies variationally in 7 out of the 8 molecular systems and depending on the GPU hardware, can complete the calculation within 2 hours depending on the size, complexity of the system, and computational hardware (Tables~\ref{tab:v100-results1} and \ref{tab:v100-results2}). The iQCC energy for system I was higher (6.8 mHa) compared to the DMRG-CI, which is noted that both non-variational perturbative methods such as PT and ILCAP-BW corrections also did not provide a lower energy \cite{ryabinkin2023_ilcap,ryabinkin2020_a_posteriori}. The iQCC energies were lower by between 0.33-1.8 mHa, however it was noted that the DMRG-CI energies could be lowered by using Fiedler ordering or by increasing the maximum bond order dimension and the number of sweeps, and was reported that for the system IX, by increasing the parameters could reduce the energy by -1.7 mHa \cite{von2021quantum}. Other reported DMRG energies for the XVIII system (64e,56o) were also higher compared to the iQCC results by 1.6 mHa \cite{physxgoogle2025}. Since the iQCC implemented here is variational, this means that for benchmarking QPE and other future quantum algorithms for the electronic structure calculation, iQCC is likely to provide a better estimate of the potential accuracy of future quantum computers.

\subsubsection{Timing and Scaling Performance}
\begin{figure}
\centering
 \includegraphics[width=\linewidth]{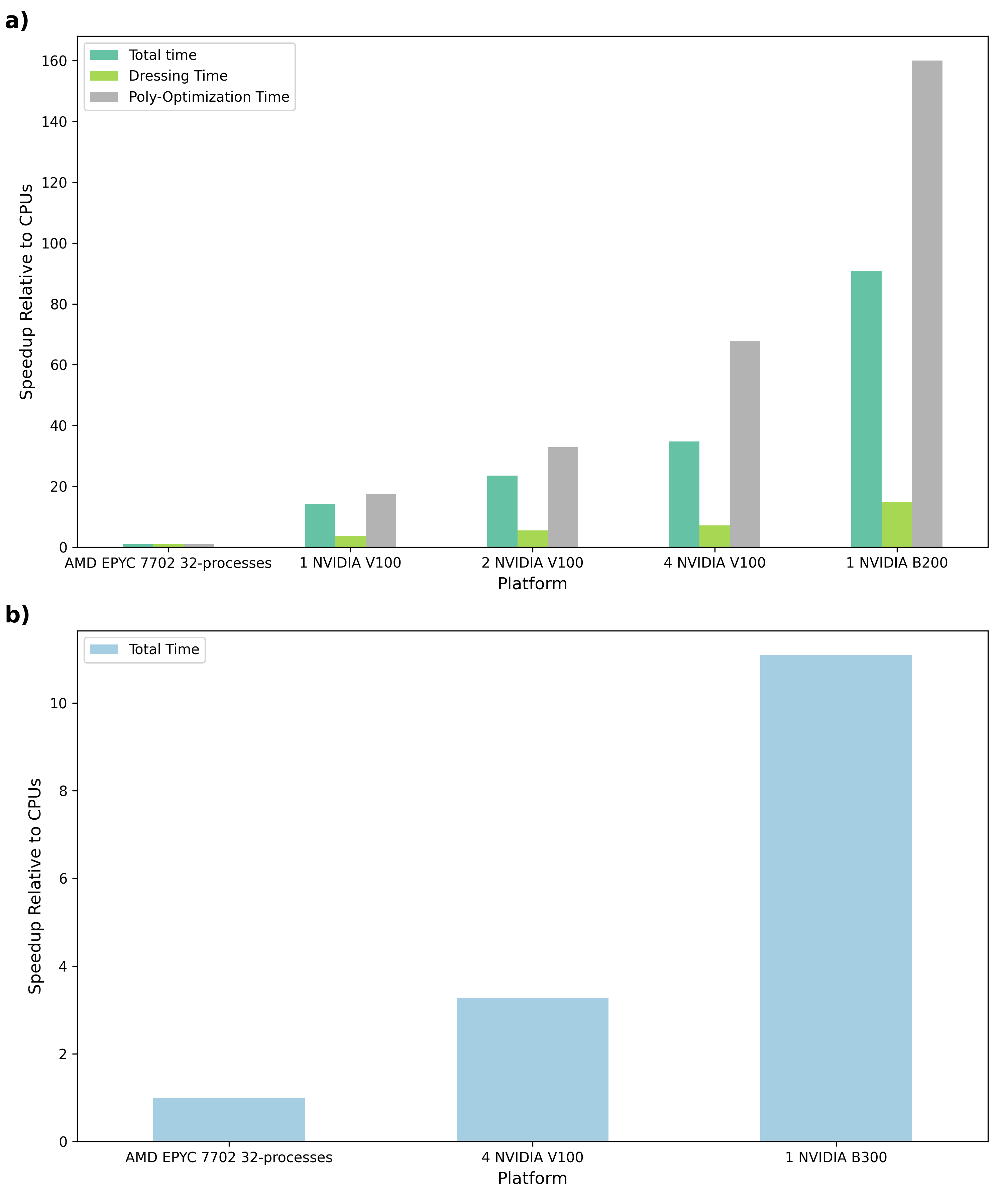}
 \caption{Relative Speedup for different compute platforms for a) XVIII CAS(64e,56o) with computational details in Table~\ref{tab:v100-results1}, and b) XVIII CAS(100e,100o) with a Hamiltonian term limit of 1,000M and 1M entanglers in the final optimization step with 100 iQCC iterations.}
 \label{fig:XVIII-scaling}
\end{figure}

The XVIII system was run on different number of V100 GPUs and on 32 CPUs to determine the speedup that the GPUs provided. Even a single V100 showed a remarkable improvement over the 32-processors on 1 AMD EPYC 7702 CPU (109 hrs vs. 7.78 hrs), even though it requires the transfer of information from the RAM to the GPU RAM due to the size of the Hamiltonian ($1.65x10^{8}$ Pauli strings). Increasing the number of V100 GPUs reduces the wallclock time of the calculation, but there are diminishing returns (Figure \ref{fig:XVIII-scaling}a). The wallclock times for the three different computational platforms using the CAS(100e,100o) XVIII system ranged from 38.78 hrs for the AMD 7702 CPU with 32 processes to 3.49 hrs on the NVIDIA Blackwell B300.

Using the B200 NVIDIA GPU which has GPU RAM of 196GB shows a 6.5 speedup over a single V100 due to it's increased GPU RAM and improved calculation speed. The GPUs provided the greatest increase in speedup specifically with respect to the final poly-optimization due to the on the fly computation of the Hessian (Figure~\ref{fig:XVIII-scaling}). The relative speedup from increasing the XVIII system from 112 qubits to 200 qubits (90x for the 112 qubit system on 1 NVIDIA B200 compared to 11x for the 200 qubit system on 1 NVIDIA B300 both normalized to AMD EPYC 7702 32 processes) is mainly driven by the number of entanglers optimized in the final step and that required number is determined by a few factors including the system and hardware specifications. Notably, for the run on a single NVIDIA B200 GPU, of the average dressing time, between 50-71\% was spent transmitting data between RAM and the GPU, which leads to an overall time of 10\%. The speedup that the B200 provided compared to the 32 CPUs was system dependent and it ranged from 90x for system XVIII to 11x for system V (Table~\ref{tab:blackwellscale}). The speed at which the iQCC can calculate the ground state energy seems to be dependent on the system size, but more importantly seems to be dependent on the nature of the system itself is attributed to the multi-configurational nature of some of the systems over others.

\begin{table*}[ht]
\centering
\caption{Select catalyst systems (II-III, V, XVIII) wall clock times across different computational hardware}
\label{tab:blackwellscale}
\begin{tabular}{lcccc}
\hline
Complex & CAS size &\shortstack{NVIDIA \\ BlackWell B200 (hrs)} & 4 NVIDIA V100 (hrs) & \shortstack{AMD 7702 32 \\processes (hrs)}\\ \hline
II-III & (70e,62o) & 4.81 & 18.34 & 126.59 \\
V & (68e,60o) & 2.02 & 3.66 & 23.16  \\
XVIII & (64e,56o) & 1.20 & 3.14 & 109.01 \\
\hline
\end{tabular}
\end{table*}

\subsection{Quantum Resource Estimates}
The iQCC algorithm is part of the VQE type quantum algorithms since it's native implementation on a universal quantum computer computes the ground state energy of the Hamiltonian by sampling the terms. Since the quantum inspired version implemented here does not require sampling, it would make sense to compare the timings and energies against future quantum computers running the QPE algorithm. For context, the largest quantum-classical electronic-structure calculation to date used a 77-qubit IBM Heron processor coupled to the Fugaku supercomputer \cite{ibm2025_sciadv_qcsc}, yet remained below the qubit counts treated here entirely on classical hardware. For simplicity, the comparison is done against a proposed Majorana quantum computer using the Azure Quantum Resource Estimator \cite{Azure_Quantum_Resource_Estimator} since the quantum computer is proposed to have favorable t-gate operation times of 100 ns. The logical qubit counts to execute the QPE were calculated here to be within the 2000-3000 range, but it is possible to use other implementations and pre-processing steps to reduce this number, and the reported best reported value as of to date is around 919 for system XVIII of CAS(64e,56o) \cite{physxgoogle2025}.
Even with recent two-orders-of-magnitude reductions in fault-tolerant QPE runtime through improved tensor factorization and active-volume compilation \cite{caesura2025_ftqc_speedup}, estimated runtimes for industrially relevant molecules remain impractical without further algorithmic-hardware co-design.

The timings for the GPU implemented iQCC are all faster than the estimated QPE time (Table~\ref{tab:QPE}) by a factor of 10-100; however, it is entirely possible that the QPU would be able to properly compute the ground state energy of complex I, implies that there is still an advantage to using the QPE algorithm on a QPU at this scale. Improvements in the native QPE algorithm have shown the potential to reduce the QPE simulation time for systems of similar size to 7.8 hrs, but this comes with the caveat of assuming $O(1)$ scaling for state preparation, which implies trading quantum computation for classical computation \cite{physxgoogle2025}. The iQCC algorithm without the multi-reference version can sometimes have issues when two states that have different spins are strongly correlated \cite{genin2022_ir,genin2025quantumadvantagechemistry}. The entangler generation step here does not preserve the exact spin and number of electrons, but the derivative and energy evaluations still consider the deviation from electron spin and number through the penalty enforcement \cite{cvqeRyabinkin2019}. Future implementations of iQCC will consider the interaction between the different states, which will also have the added benefit of computing the additional states as well\cite{multi-stateiqcc2026}.

\begin{table*}[ht]
\centering
\caption{QPE Resource Estimates based on a Majorana Quantum Computer using the Floquet error correction to compute the ground state energies to within 1 mHa using default settings. \cite{Azure_Quantum_Resource_Estimator,microsoft_majorana2025}}
\label{tab:QPE}
\begin{tabular}{lcccc}
\hline
Complex & System Qubits & Logical Qubits & Physical Qubits & \shortstack{QPE Run \\Time (hrs)} \\ \hline
I & 104 & 2479 & 904,396 & 76.36 \\
II & 124 & 2993 & 1,394,244 & 292.51   \\
II-III & 130 & 3111 & 1,440,028 & 362.66  \\
V & 120 & 2909 & 1,361,652 &  253.66 \\
VIII & 130 & 3111 & 1,440,028 & 378.87 \\
VIII-IX & 118 & 2871 & 13,46,908 & 261.09 \\
IX & 124 & 2987 & 1,408,556 & 270.61 \\
XVIII & 112 & 2740 & 1,296,080 & 199.63 \\
\hline
\end{tabular}
\end{table*}

\section{Discussion}
The pursuit of quantum advantage has long focused on the `50-qubit' horizon,
a limit derived largely from the memory constraints of state-vector simulation
\cite{lanes2025frameworkquantumadvantage}. While recent advances in sparse wave function representations \cite{steiger2024_sparse} and tensor methods like TruSTS \cite{miller2026_trusts} have extended classical emulation of VQE circuits to the 64--92 qubit range, these approaches remain fundamentally constrained by the scaling of the state-vector or bond-dimension. This still falls far below the 100--124 qubit regime accessed here through operator sparsity. Similarly, state-of-the-art mixed-precision DMRG
on NVIDIA Blackwell GPUs achieves chemical accuracy for active spaces up to
CAS(113,76) \cite{brower2025_blackwell_dmrg}, but its performance remains
governed by bond-dimension scaling, which grows steeply for strongly correlated
transition-metal systems. More broadly, GPU-accelerated 2D tensor networks have demonstrated that careful algorithmic design can defeat the conventional 50-qubit intractability assumption even for quantum circuit simulation \cite{rudolph2025_2dtn}. However, our results with Parallel iQCC suggest that for electronic structure, this boundary has been significantly underestimated. By shifting the bottleneck from memory (state vector) to operator sparsity (iQCC), we demonstrate that systems in the 100–124 qubit regime—such as the Ruthenium catalysts studied here—remain well within the reach of modest GPU clusters. This effectively 'de-quantizes' a significant portion of the NISQ roadmap. Notably, the transition-metal catalysts studied here belong to the ``class-2'' electronic structures identified as the most challenging for classical computation and the presumed first targets for quantum advantage \cite{morchen2024_classification}; our results show that even these systems yield to GPU-accelerated polynomial iQCC. It implies that true quantum advantage for chemistry will not be found in merely scaling qubit counts to 100, as demonstrated in recent hardware milestones \cite{google2024_errorcorr}, but rather in accessing entanglement complexities that elude even the polynomial compression of iQCC. Until quantum hardware can surpass the precision and time-to-solution of methods like GPU-accelerated iQCC ($<5$ hours), classical algorithms remain the most practical tool for high-accuracy chemical discovery. The most ambitious quantum-classical effort to date required coupling a 77-qubit IBM Heron processor with the Fugaku supercomputer \cite{ibm2025_sciadv_qcsc}—a massive infrastructure investment that our fully classical GPU-parallel iQCC surpasses in both qubit count and time-to-solution.
Compounding this, even state-of-the-art NISQ implementations of VQE fail to achieve quantitatively descriptive energies due to hardware noise \cite{carreras2025_vqe_limitations}, further underscoring that classical methods remain indispensable for near-term chemical discovery. Even on the fault-tolerant horizon, recent advances that reduce QPE runtimes by two orders of magnitude still leave estimated wall times impractical for industrial chemistry workflows \cite{caesura2025_ftqc_speedup}.

The iQCC method running on GPUs using a single reference has shown the ability to out perform reference DMRG-CI energies in seven out of eight Ruthenium catalyst systems, but struggles when the system has a multi-reference character. As shown, running the same parameters on the same system can result in slight variations between simulations, which is due to the entangler selection process. During the entangler selection for dressing, some entanglers will have the same estimated gradient, but if the cut off for the number of entanglers dressed, the code can make random decisions of which one is prioritized. The optimization takes a slightly different path and thus can result in small discrepancies in the final energy.

The GPU accelerated iQCC algorithm has significant implications for the broader chemistry and quantum community. The two implications are 1) what benefit can a variational algorithm that can be solved in a few hours, and 2) what does this mean for quantum supremacy? A major suggestion for the use of quantum algorithms is for a variational bounded geometry optimization and transition state search \cite{FemocoPNAS2017,von2021quantum}. It is well known that non-variational (such as CCSD) and single reference methods (such as CISD) can struggle with accurately representing the potential energy surface of a chemical reaction. With respect to quantum supremacy, that fully remains to be seen. There have been conjectures and theoretical challenges to quantum advantage in chemistry \cite{Lee2023,zhai2026classicalsolutionfemocofactormodel}, but they do not provide computational timing statistics and as has been shown elsewhere, that while near linear scaling coupled cluster methods are available, there are concrete examples of chemical systems where coupled cluster methods show poor experimental prediction\cite{genin2025quantumadvantagechemistry}. Here, we have presented the exact ground state energy requirements and timings required for a quantum computer to surpass what can be done classically without the need of a supercomputer cluster.

Our ability to efficiently simulate systems of 100--200 qubits with parallel iQCC provides numerical evidence for the recently proposed connection between the absence of barren plateaus and classical simulability \cite{barrenplateau2025}. The BP phenomenon arises as a curse of dimensionality in highly expressive parameterized circuits, rendering the optimization landscape exponentially flat and gradient information vanishingly small \cite{Larocca_2025}. iQCC naturally evades this problem by construction: at each iteration, entanglers are selected exclusively from the Direct Interaction Space (DIS), which guarantees non-zero energy gradients by design, and the expansion of the Hamiltonian is constrained by truncation\cite{ryabinkin2019_iqcc}. However, as posited by Cerezo \textit{et al.} \cite{barrenplateau2025}, the very mathematical structure that restricts the variational evolution to avoid barren plateaus also confines it to an operator subspace amenable to efficient classical emulation. Our GPU-accelerated solver exploits precisely this confinement, tracking the accessible subspace through polynomial approximation rather than exponential state-vector expansion. This reinforces the conclusion that simply avoiding barren plateaus is insufficient for quantum advantage; true advantage will require algorithms capable of navigating highly entangled, non-classically-simulable regions of Hilbert space without falling victim to BPs---a challenge that remains open.




\section{Acknowledgments}
The authors acknowledge the funding provided by Innovation Solutions Canada project \#202208-F0033-C00003 and Next Generation Manufacturing Grant \#14234. The authors also want to thank Jin-Sung Kim and Tom Lubowe for their support in providing access to NVIDIA Blackwell B200 and B300.

\appendix

\section{Overview of the Iterative Qubit Coupled Cluster (iQCC) Method}
\label{appendix:iqcc_overview}

The iterative qubit coupled cluster (iQCC) method is a quantum-classical hybrid algorithm designed to estimate the ground-state energy of electronic systems using fixed-depth quantum circuits. This appendix provides a concise summary of the iQCC methodology and mathematical foundations, focusing exclusively on its single-reference implementation.

\subsection*{Qubit Hamiltonian and Reference State}

The electronic structure problem begins with the second-quantized Hamiltonian:
\begin{equation}
\hat{H}_e = \sum_{ij} h_{ij} \hat{a}_i^\dagger \hat{a}_j + \frac{1}{2} \sum_{ijkl} g_{ijkl} \hat{a}_i^\dagger \hat{a}_j^\dagger \hat{a}_l \hat{a}_k,
\end{equation}
where $\hat{a}_i^\dagger$ and $\hat{a}_j$ are fermionic creation and annihilation operators, and $h_{ij}$, $g_{ijkl}$ are one- and two-electron integrals.

A fermion-to-qubit transformation (e.g., Jordan--Wigner or Bravyi--Kitaev) maps this to a qubit Hamiltonian:
\begin{equation}
\hat{H} = \sum_k C_k \hat{P}_k,
\end{equation}
where each $\hat{P}_k$ is a Pauli word---a tensor product of $\hat{x}_j, \hat{y}_j, \hat{z}_j$ on individual qubits.

The reference wavefunction is taken to be a product of single-qubit states:
\begin{equation}
|\Omega\rangle = \bigotimes_{j=1}^n \left( \cos \frac{\theta_j}{2} |\uparrow\rangle_j + e^{i \phi_j} \sin \frac{\theta_j}{2} |\downarrow\rangle_j \right),
\end{equation}
with Bloch angles $\{\theta_j, \phi_j\}$ as variational parameters. This is referred to as the qubit mean-field (QMF) state.

\subsection*{QCC Ansatz and Energy Gradient}

The QCC ansatz applies a unitary transformation generated by Pauli operators:
\begin{equation}
\hat{U}(\boldsymbol{\tau}) = \prod_k \exp\left(-i \frac{\tau_k}{2} \hat{P}_k\right),
\end{equation}
and the variational energy becomes:
\begin{equation}
E(\boldsymbol{\tau}, \Omega) = \langle \Omega | \hat{U}^\dagger(\boldsymbol{\tau}) \hat{H} \hat{U}(\boldsymbol{\tau}) | \Omega \rangle.
\end{equation}

To construct the ansatz iteratively, one selects generators $\hat{P}_\alpha$ with the highest energy gradients evaluated at $\tau = 0$:
\begin{equation}
\left.\frac{\partial E}{\partial \tau_\alpha}\right|_{\tau=0} = -\frac{i}{2} \langle \Omega | [\hat{H}, \hat{P}_\alpha] | \Omega \rangle = \text{Im} \langle \Omega | \hat{H} \hat{P}_\alpha | \Omega \rangle.
\end{equation}

\subsection*{Direct Interaction Space (DIS)}

To efficiently select high-impact Pauli operators, iQCC defines the \emph{Direct Interaction Space} (DIS), a subset of Pauli words with non-zero gradients. Only Pauli operators satisfying the following two conditions can belong to the DIS:

\begin{enumerate}
  \item \textbf{Odd $\hat{y}$ Parity:} A non-zero imaginary contribution requires that the operator $\hat{P}_\alpha$ has an odd number of $\hat{y}$ operators.

  \item \textbf{Matching Flip Set:} $\hat{P}_\alpha$ must have the same set of qubit flips (i.e., positions of $\hat{x}$ or $\hat{y}$ terms) as some term $\hat{P}_k$ in $\hat{H}$. Denote this flip set as $F(\hat{P}) = \{j: \hat{P}_j \in \{\hat{x}_j, \hat{y}_j\}\}$.
\end{enumerate}

These constraints can be derived as follows. Let the Hamiltonian be decomposed as:
\begin{equation}
\hat{H} = \sum_i \left( \sum_j \eta_j^{(i)} \hat{Z}_j^{(i)} \right) \hat{X}_i,
\end{equation}
where $\hat{X}_i$ is a product of $\hat{x}$ operators, and $\hat{Z}_j^{(i)}$ are tensor products of $\hat{z}$ operators. Acting on the QMF state $|\Omega\rangle$, which is an eigenstate of $\hat{z}_j$, gives:
\begin{align}
\text{Im} \langle \Omega | \hat{H} \hat{P}_\alpha | \Omega \rangle &\propto \text{Im}(\theta_\alpha) \sum_i \left( \sum_j \eta_j^{(i)} \lambda_j^{(i)} \right) \delta_{\mu^{(\alpha)}, \mu^{(i)}},
\end{align}
where $\theta_\alpha$ is a complex prefactor ensuring $\hat{P}_\alpha$ is Hermitian, and $\mu^{(i)}$ is the binary vector encoding the positions of $\hat{x}$ operators in $\hat{X}_i$. The Kronecker delta $\delta_{\mu^{(\alpha)}, \mu^{(i)}}$ enforces that only terms with matching flip sets contribute. Further, $\text{Im}(\theta_\alpha)$ vanishes unless the number of $\hat{y}$ operators in $\hat{P}_\alpha$ is odd.

Hence, both conditions above are necessary for $\hat{P}_\alpha$ to contribute to the gradient.

\subsection*{Iterative Procedure}

The iQCC algorithm proceeds iteratively as follows:
\begin{enumerate}
  \item Evaluate the energy gradient of candidate Pauli operators in the DIS.
  \item Select the top $N_g$ generators with largest gradients.
  \item Optimize the energy with respect to their amplitudes and the QMF parameters.
  \item Dress the Hamiltonian: $\hat{H} \rightarrow \hat{U}^\dagger \hat{H} \hat{U}$.
  \item Optionally compress small terms in the dressed Hamiltonian.
  \item Repeat until convergence.
\end{enumerate}

This procedure enables systematic convergence toward the ground state using a fixed-depth circuit at each iteration, making it well-suited for NISQ-era quantum devices.


\section{Bit-wise Partitioning in iQCC}
\label{appendix:bitwise}

\subsection{State of the Problem}

A central bottleneck of the iterative qubit coupled cluster (iQCC) algorithm
is the exponential growth of Pauli words in the Hamiltonian under successive
dressing transformations.
After several iterations, the number of terms becomes overwhelmingly large,
quickly exhausting available RAM and rendering single--node execution infeasible.

Distributing the workload across multiple CPUs or GPUs is therefore essential.
However, naïve parallelization faces two fundamental challenges:

\begin{enumerate}
    \item \textbf{All-to-all communication.}
          If every node must synchronize its terms with all others at each dressing step,
          communication costs dominate, preventing scalability.
    \item \textbf{Duplication of Pauli words.}
          If communication is avoided, each node eventually reproduces the entire Hamiltonian,
          so all nodes perform the same work, reducing performance back to that of a single CPU.
\end{enumerate}

\subsection{Bit-wise Partitioning as a Solution}

Bit-wise partitioning overcomes both issues simultaneously.
Each Pauli term is encoded in binary form,
and a fixed subset of bit positions is used to assign the term to a specific node.
For $m$ partitioning bits, the Hamiltonian is divided into $2^m$ disjoint subsets,
each uniquely mapped to an MPI rank or GPU device.
This guarantees:

\begin{itemize}
    \item Every Pauli term resides on exactly one node (no duplication).
    \item Communication is required only when entanglers flip partitioning bits.
    \item When communication occurs, it is restricted to fixed \emph{pairs} of nodes,
          eliminating the need for all-to-all exchange.
\end{itemize}

The iQCC algorithm has two main computational steps where partitioning plays a decisive role:
\textbf{(i) dressing of Hamiltonian terms} and \textbf{(ii) energy evaluation (or Hessian/gradient calculation in the polynomial optimizer)}.
We now describe each in turn.

\subsection{Application to Dressing}

Dressing a Hamiltonian with an entangler involves XOR operations between Pauli words.
Bit-wise partitioning modifies this step as follows:

\begin{enumerate}
    \item Each CPU stores only the Pauli terms belonging to its partition.
    \item For an entangler that does not touch partitioning bits,
          all \emph{new terms produced during dressing} remain in the same partition,
          requiring no communication.
    \item If the entangler flips one or more partitioning bits,
          the new terms must be moved to their correct partitions so they can be combined
          with the existing terms in those partitions.
          \begin{enumerate}
              \item Each CPU determines the destination rank for its new terms.
              \item Communication occurs only between paired CPUs whose partitions differ
                    in the flipped bits.
              \item If the new term is unique in its destination partition, it is added.
                    If it already exists, its amplitude is combined with the existing amplitude.
          \end{enumerate}
    \item This pairwise communication represents the \emph{minimum possible overhead in terms of data transfer},
          as the communications are the possible minimum required to ensure correctness.
\end{enumerate}

\subsection{Application to Energy and Gradient Evaluation}

Energy evaluation (and Hessian/gradient construction in the polynomial optimizer)
is naturally parallel under bit-wise partitioning:

\begin{enumerate}
    \item Each CPU evaluates its local contribution to the total energy
          using only its assigned Hamiltonian terms.
    \item In the case of polynomial optimization:
          \begin{itemize}
              \item Each CPU constructs its \emph{own full Hessian matrix},
                    but each element $h_{ij}$ corresponds only to its local contribution.
              \item The global Hessian is obtained as
                    \begin{equation}
                        h_{ij}^{\text{total}} = \sum_{\text{CPU}} h_{ij}^{\text{(local)}} .
                    \end{equation}
          \end{itemize}
    \item Similarly, gradients and energies are additive: each CPU computes its portion,
          and a designated master node (or an MPI reduction) sums them into the final result.
\end{enumerate}

Since all these quantities are linear in the Hamiltonian,
this step is embarrassingly parallel and requires only one global reduction.

\subsection{Summary}

Bit-wise partitioning is the enabling technique that allows iQCC to scale beyond
the memory and communication limits of naïve parallelization.
It ensures that:

\begin{itemize}
    \item Pauli terms are distributed without duplication,
    \item Dressing requires only pairwise communication when necessary, and
    \item Energy and Hessian/gradient evaluations scale trivially across many nodes.
\end{itemize}

This approach allows the exponential growth of iQCC Hamiltonians
to be handled on modern multi-CPU and multi-GPU architectures,
making simulations with hundreds of qubits computationally tractable.


\section{Load Balancing in Parallel iQCC}
\label{appendix:loadbalancing}

\subsection{State of the Problem}

Under bit-wise partitioning, Pauli words are distributed into partitions according
to selected bits of their binary representation.
This distribution is not uniform: some partitions naturally contain more terms than others.
At initialization, partitioning bits are chosen to make the distribution as flat as possible,
but as iQCC iterations proceed, Hamiltonian dressing generates new terms unevenly across partitions.
Without correction, some CPUs accumulate many more Pauli words than others,
leading to severe load imbalance and poor parallel efficiency.

\subsection{General Approach}

To mitigate imbalance we employ two complementary ideas:

\begin{enumerate}
    \item \textbf{Overpartitioning.}
          The Hamiltonian is divided into more partitions than the number of CPUs.
          Each CPU owns a set of partitions rather than a single one.
          This finer granularity enables redistribution when imbalance develops.
    \item \textbf{Dynamic reassignment.}
          After dressing steps, the number of terms in each partition is measured.
          If imbalance is detected, whole partitions are reassigned from overloaded CPUs
          to underloaded ones.
          This avoids splitting partitions and preserves the logic of bit-wise partitioning.
\end{enumerate}

\subsection{Communication During Dressing}

Bit-wise partitioning ensures that new terms produced during dressing are routed
to their correct partitions:

\begin{enumerate}
    \item If the entangler does not touch partitioning bits,
          all new terms remain local to the CPU that owns the partition.
    \item If the entangler flips partitioning bits,
          the new terms must be transferred to their destination partitions so they can be combined
          with the existing terms there:
          \begin{enumerate}
              \item Each CPU determines the destination partition for its new terms.
              \item Communication occurs only with the CPU(s) that own those partitions.
              \item In the destination partition, if a new term is unique it is added,
                    otherwise its amplitude is combined with the amplitude of the existing term.
          \end{enumerate}
\end{enumerate}

When each CPU owns exactly one partition, communication is strictly pairwise.
When load balancing assigns multiple partitions to a CPU, that CPU may need to
communicate with several peers during a dressing step.
This is no longer the minimum possible communication, but it remains far better than
all-to-all exchange, because each transfer is restricted to the specific partitions affected
by the entangler.

\subsection{Energy and Gradient Evaluation}

For energy evaluation (and gradient/Hessian construction in the polynomial optimizer),
linearity makes the calculation embarrassingly parallel:

\begin{enumerate}
    \item Each CPU computes its local contribution to the total energy using the partitions it owns.
    \item In polynomial optimization, each CPU also computes its full local gradient and Hessian matrices,
          where each element is only a partial contribution.
    \item A global reduction aggregates results into the final quantities:
          \begin{equation}
              E = \sum_{\text{CPU}} E^{\text{(local)}}, \quad
              g_i = \sum_{\text{CPU}} g_i^{\text{(local)}}, \quad
              h_{ij} = \sum_{\text{CPU}} h_{ij}^{\text{(local)}} .
          \end{equation}
\end{enumerate}

Rebalancing changes only the ownership of partitions; the aggregation logic remains unchanged.

\subsection{Summary}

Load balancing maintains high utilization across CPUs in parallel iQCC.
By reassigning entire partitions when imbalance grows, it prevents idle processors
and avoids duplication of Pauli words.
Although multiple partitions per CPU can increase the number of peers contacted in dressing steps,
communication is still confined to partition pairs and remains well below all-to-all requirements.
This strategy preserves the scalability of iQCC on large multi-CPU and multi-GPU platforms.


\section{Sortless Dressing in iQCC}
\label{appendix:sortless}

\subsection{State of the Problem}

Each dressing step in iQCC applies an entangler to the Hamiltonian $H$, producing
a set of new Pauli words $I$. The dressed Hamiltonian is
\begin{equation}
    D = H + I ,
\end{equation}
where $+$ denotes merging the two sets: new words are inserted if unique, and if a word
already exists in $H$, its coefficient is updated by adding the new contribution.
Because $|I|$ is typically $O(|H|)$ and $I$ is generated in an unsorted order, the
conventional "sort $I$ then merge" procedure becomes a bottleneck as $H$ grows.

\subsection{Key Idea of Sortless Dressing}

The goal is to eliminate the explicit sorting of $I$ by leveraging the fact that $H$ is
kept globally sorted and that the entangler's action preserves order \emph{within}
entangler-defined partitions.

\begin{enumerate}
    \item \textbf{Partition $H$.} Divide $H$ into subsets $\{H_i\}$ according to the set bits
          of the entangler. If the entangler acts on $n$ qubits, this yields $2^n$ partitions.
    \item \textbf{Local dressing.} Apply the entangler to each partition $H_i$, generating a
          corresponding set of new terms $I_i$. Because $H_i$ is sorted and the entangler
          induces a fixed bitwise mapping, the resulting $I_i$ is also sorted.
    \item \textbf{Merge per partition.} For each $i$, merge $I_i$ into $H_i$ directly:
          insert a term if it is new, otherwise combine (add) coefficients for duplicates.
          This produces dressed, still-sorted partitions $D_i = H_i + I_i$.
    \item \textbf{Reconstruct $D$ (globally sorted).} Perform a deterministic multiway
          merge over the sorted dressed partitions $\{D_i\}$ using the same global
          ordering used for $H$. The result is a single \emph{globally sorted} dressed
          Hamiltonian $D$, which becomes the next iteration's input $H$.
\end{enumerate}

\subsection{Advantages}

Sortless dressing replaces a costly global sort of the raw production list $I$ with a
partition-aware workflow that preserves order locally and restores it globally by a final
multiway merge. By partitioning $H$ with respect to the entangler's set bits, the
entangler's action maps each sorted $H_i$ into a sorted $I_i$, so duplicate elimination
(coefficient combination) is handled naturally during the small, local merges $H_i \leftarrow H_i + I_i$.
A concluding $k$-way merge over the already-sorted $\{D_i\}$ reconstitutes a single,
globally sorted Hamiltonian $D$ ready for the next iQCC iteration. This organization
keeps the key invariants—uniqueness and sortedness—without an explicit global sort of $I$
and avoids the unnecessary overhead associated with handling $I$ as one monolithic list.



\bibliography{references}

@article{steiger2024_sparse,
  title={Sparse Simulation of VQE Circuits},
  author={Steiger, Damian S and H{\"a}ner, Thomas and Genin, Scott N and Katzgraber, Helmut G},
  journal={arXiv preprint arXiv:2404.10047},
  year={2024}
}

@article{peruzzo2014_vqe,
  author    = {Peruzzo, Alberto and McClean, Jarrod and Shadbolt, Peter and Yung, Man-Hong and Zhou, Xiao-Qi and Love, Peter J. and Aspuru-Guzik, Al{\'a}n and O’Brien, Jeremy L.},
  title     = {A variational eigenvalue solver on a photonic quantum processor},
  journal   = {Nature Communications},
  volume    = {5},
  year      = {2014},
  pages     = {4213},
  doi       = {10.1038/ncomms5213},
}

@article{ryabinkin2018_qcc,
  author    = {Ryabinkin, Ilya G. and Yen, Tzu-Ching and Genin, Scott N. and Izmaylov, Artur F.},
  title     = {Qubit coupled-cluster method: A systematic approach to quantum chemistry on a quantum computer},
  journal   = {Journal of Chemical Theory and Computation},
  volume    = {14},
  number    = {12},
  year      = {2019},
  pages     = {6317--6326},
  doi       = {10.1021/acs.jctc.9b00903},
}

@article{ryabinkin2019_iqcc,
  title = {Iterative Qubit Coupled Cluster Approach with Efficient Screening of Generators},
  author = {Ryabinkin, Ilya G. and Lang, Robert A. and Genin, Scott N. and Izmaylov, Artur F.},
  journal = {Journal of Chemical Theory and Computation},
  year = {2020},
  volume = {16},
  number = {2},
  pages = {1055--1063},
  doi = {10.1021/acs.jctc.9b00745},
}

@article{ryabinkin2023_ilcap,
  author    = {Ryabinkin, Ilya G. and Jena, Andrew J. and Genin, Scott N.},
  title     = {Efficient construction of involutory linear combinations of anticommuting Pauli generators for large-scale iterative qubit coupled cluster calculations},
  journal   = {Journal of Chemical Theory and Computation},
  volume    = {19},
  number    = {6},
  year      = {2023},
  pages     = {1722--1733},
  doi       = {10.1021/acs.jctc.2c01115},
}

@article{ryabinkin2020_a_posteriori,
  title={A posteriori corrections to the iterative qubit coupled cluster method to minimize the use of quantum resources in large-scale calculations},
  author={Ryabinkin, Ilya G and Izmaylov, Artur F and Genin, Scott N},
  journal={Quantum Science and Technology},
  volume={6},
  number={2},
  pages={024012},
  year={2021},
  publisher={IOP Publishing}
}

@article{ryabinkin2025_optimization,
  title = {Optimization of the Qubit Coupled Cluster Ansatz on Classical Computers},
  author = {Ryabinkin, Ilya G. and Hosseini Jenab, Seyyed Mehdi and Genin, Scott N.},
  journal = {Journal of Chemical Theory and Computation},
  year = {2025},
  volume = {11},
  number = {8},
  pages = {3616--3625},
  doi = {10.1021/acs.jctc.5c00345},
}

@article{genin2022_ir,
  title = {Estimating Phosphorescent Emission Energies in {Ir(III)} Complexes using Large-Scale Quantum Computing Simulations},
  author = {Genin, Scott N. and Ryabinkin, Ilya G. and Paisley, Nathan R. and Whelan, Sarah O. and Helander, Michael G. and Hudson, Zachary M.},
  journal = {Angewandte Chemie International Edition},
  year = {2022},
  volume = {61},
  pages = {e202116175},
  doi = {10.1002/anie.202116175},
}

@misc{quantinuum2025,
  author       = {Quantinuum},
  title        = {Quantinuum Helios: Delivering Fault-Tolerant Quantum Computing by 2025},
  year         = {2025},
  note         = {Technical Perspective / White Paper},
  url          = {https://www.quantinuum.com},
}

@misc{aws2023_sparse,
  author       = {Microsoft Quantum Team},
  title        = {Testing large quantum algorithms using sparse simulation},
  year         = {2023},
  howpublished = {Microsoft Quantum Blog},
  url          = {https://quantum.microsoft.com/en-us/insights/blogs/qsharp/testing-large-quantum-algorithms-using-sparse-simulation},
}

@article{google2023_tensor,
  author    = {Bidzhiev, K. and Grijalva, S. and others},
  title     = {Cloud on-demand emulation of quantum dynamics with tensor networks},
  journal   = {arXiv preprint},
  year      = {2023},
  eprint    = {2302.05253},
  archivePrefix = {arXiv},
  note      = {Preprint},
  url       = {https://arxiv.org/abs/2302.05253},
}

@article{npj2023_nasc,
  author    = {Shang, H. and Jiang, M. and Li, Z. and others},
  title     = {Towards practical and massively parallel quantum simulation of chemistry with tensor networks},
  journal   = {npj Quantum Information},
  volume    = {9},
  number    = {1},
  year      = {2023},
  pages     = {41},
  doi       = {10.1038/s41534-023-00696-7},
}

@article{google2019_supremacy,
  author    = {Arute, F. and Arya, K. and Babbush, R. and et al.},
  title     = {Quantum supremacy using a programmable superconducting processor},
  journal   = {Nature},
  volume    = {574},
  year      = {2019},
  pages     = {505--510},
  doi       = {10.1038/s41586-019-1666-5},
}

@article{google2024_errorcorr,
  title = {Quantum error correction below the surface code threshold},
  author = {Google Quantum AI and Collaborators},
  journal = {Nature},
  year = {2025},
  volume = {638},
  number = {8052},
  pages = {920--926},
  publisher = {Nature Publishing Group UK London},
}

@article{ibm2025_sciadv_qcsc,
  title={Chemistry beyond the scale of exact diagonalization on a quantum-centric supercomputer},
  author={Robledo-Moreno, Javier and Motta, Mario and Haas, Holger and Javadi-Abhari, Ali and Jurcevic, Petar and Kirby, William and Martiel, Simon and Sharma, Kunal and Sharma, Sandeep and Shirakawa, Tomonori and others},
  journal={Science Advances},
  volume={11},
  number={25},
  pages={eadu9991},
  year={2025},
  publisher={American Association for the Advancement of Science}
}

@article{von2021quantum,
  author    = {von Burg, Vera and Low, Guang Hao and H\"aner, Thomas and Steiger, Damian S. and Reiher, Markus and R\"otteler, Martin and Troyer, Matthias},
  title     = {Quantum computing enhanced computational catalysis},
  journal   = {Physical Review Research},
  volume    = {3},
  number    = {3},
  year      = {2021},
  pages     = {033055},
  doi       = {10.1103/PhysRevResearch.3.033055},
}

@article{abrams1999_qpe_prl,
  author    = {Abrams, Daniel S. and Lloyd, Seth},
  title     = {Quantum algorithm providing exponential speed increases for finding eigenvalues and eigenvectors},
  journal   = {Physical Review Letters},
  volume    = {83},
  number    = {24},
  year      = {1999},
  pages     = {5162--5165},
  doi       = {10.1103/PhysRevLett.83.5162},
}

@article{shang2023massive,
  title={Towards practical and massively parallel quantum computing emulation for quantum chemistry},
  author={Shang, Honghui and Fan, Yi and Shen, Li and Guo, Chu and Liu, Jie and Duan, Xiaohui and Li, Fang and Li, Zhenyu},
  journal={NPJ quantum information},
  volume={9},
  number={1},
  pages={33},
  year={2023},
  publisher={Nature Publishing Group UK London}
}

@article{mccaskey2019quantum,
  title={Quantum chemistry as a benchmark for near-term quantum computers},
  author={McCaskey, Alexander J and Parks, Zachary P and Jakowski, Jacek and Moore, Shirley V and Morris, Titus D and Humble, Travis S and Pooser, Raphael C},
  journal={npj Quantum Information},
  volume={5},
  number={1},
  pages={99},
  year={2019},
  publisher={Nature Publishing Group UK London}
}

@article{mihalikova2022best,
  author    = {Mih{\'a}likov{\'a}, Ivana and Fri{\'a}k, Martin and Pivoluska, Matej and Plesch, Martin and Saip, Martin and {\v{S}}ob, Mojm{\'\i}r},
  title     = {Best-practice aspects of quantum-computer calculations: A case study of the hydrogen molecule},
  journal   = {Molecules},
  volume    = {27},
  number    = {3},
  year      = {2022},
  pages     = {597},
  doi       = {10.3390/molecules27030597},
}

@Article{GHGcatalyst2015,
author ="Wesselbaum, Sebastian and Moha, Verena and Meuresch, Markus and Brosinski, Sandra and Thenert, Katharina M. and Kothe, Jens and Stein, Thorsten vom and Englert, Ulli and Hölscher, Markus and Klankermayer, Jürgen and Leitner, Walter",
title  ="Hydrogenation of carbon dioxide to methanol using a homogeneous ruthenium–Triphos catalyst: from mechanistic investigations to multiphase catalysis",
journal  ="Chem. Sci.",
year  ="2015",
volume  ="6",
issue  ="1",
pages  ="693-704",
publisher  ="The Royal Society of Chemistry",
doi  ="10.1039/C4SC02087A",
url  ="http://dx.doi.org/10.1039/C4SC02087A"
}

@misc{lanes2025frameworkquantumadvantage,
      title={A Framework for Quantum Advantage}, 
      author={Olivia Lanes and Mourad Beji and Antonio D. Corcoles and Constantin Dalyac and Jay M. Gambetta and Loic Henriet and Ali Javadi-Abhari and Abhinav Kandala and Antonio Mezzacapo and Christopher Porter and Sarah Sheldon and John Watrous and Christa Zoufal and Alexandre Dauphin and Borja Peropadre},
      year={2025},
      eprint={2506.20658},
      archivePrefix={arXiv},
      primaryClass={quant-ph},
      url={https://arxiv.org/abs/2506.20658}, 
}

@article{ryabinkin-qmf-2018,
title={Relation between fermionic and qubit mean fields in the electronic structure problem},
author={Ilya G. Ryabinkin and Scott N. Genin and Artur F. Izmaylov},
journal={J. Chem. Phys},
volume={149},
number={21},
pages={214105},
year={2018},
doi={https://doi.org/10.1063/1.5055357}

}

@article{microsoft_majorana2025,
title={Interferometric single-shot parity measurement in InAs–Al hybrid devices},
author={Morteza Aghaee and others},
journal={Nature},
year={2025},
volume={638},
pages={651–655}

}

@inproceedings{Azure_Quantum_Resource_Estimator,
   author = {van Dam, Wim and Mykhailova, Mariia and Soeken, Mathias},
   title = {{Using Azure Quantum Resource Estimator for Assessing Performance of Fault Tolerant Quantum Computation}},
   year = {2023},
   isbn = {9798400707858},
   publisher = {Association for Computing Machinery},
   address = {New York, NY, USA},
   url = {https://doi.org/10.1145/3624062.3624211},
   doi = {10.1145/3624062.3624211},
   booktitle = {Proceedings of the SC '23 Workshops of The International Conference on High Performance Computing, Network, Storage, and Analysis},
   pages = {1414<E2><80><93>1419},
   numpages = {6},
   series = {SC-W '23} 
}

@article{barrenplateau2025,
    author = {M. Cerezo and Martin Larocca and Diego García-Martín and N. L. Diaz and Paolo Braccia and Enrico Fontana and Manuel S. Rudolph and Pablo Bermejo and Aroosa Ijaz and Supanut Thanasilp and Eric R. Anschuetz and Zoë Holmes},
    title = {Does provable absence of barren plateaus imply classical simulability?},
    journal = {Nature Communications},
    volume = {16},
    pages = {7907},
    year = {2025}
}

@article{FemocoPNAS2017,
author = {Markus Reiher  and Nathan Wiebe  and Krysta M. Svore  and Dave Wecker  and Matthias Troyer },
title = {Elucidating reaction mechanisms on quantum computers},
journal = {Proceedings of the National Academy of Sciences},
volume = {114},
number = {29},
pages = {7555-7560},
year = {2017},
doi = {10.1073/pnas.1619152114},
URL = {https://www.pnas.org/doi/abs/10.1073/pnas.1619152114},
eprint = {https://www.pnas.org/doi/pdf/10.1073/pnas.1619152114}
}

@misc{genin2025quantumadvantagechemistry,
      title={Towards Quantum Advantage in Chemistry}, 
      author={Scott N. Genin and Ohyun Kwon and Seyyed Mehdi Hosseini Jenab and Seon-Jeong Lim and Taehyung Kim and Tae-Gon Kim and Rami Gherib and Angela F. Harper and Ilya G. Ryabinkin and Michael G. Helander},
      year={2025},
      eprint={2512.13657},
      archivePrefix={arXiv},
      primaryClass={physics.chem-ph},
      url={https://arxiv.org/abs/2512.13657}, 
}

@article{cvqeRyabinkin2019,
author = {Ryabinkin, Ilya G. and Genin, Scott N. and Izmaylov, Artur F.},
title = {Constrained Variational Quantum Eigensolver: Quantum Computer Search Engine in the {F}ock Space},
journal = {Journal of Chemical Theory and Computation},
volume = {15},
number = {1},
pages = {249-255},
year = {2019},
doi = {10.1021/acs.jctc.8b00943},
    note ={PMID: 30512959},
URL = { 
    
        https://doi.org/10.1021/acs.jctc.8b00943
 
    

},
eprint = { 
    
        https://doi.org/10.1021/acs.jctc.8b00943
    
    

}

}

@article{Lee2023,
  title = {Evaluating the evidence for exponential quantum advantage in ground‐state quantum chemistry},
  author = {Lee, Seunghoon and Lee, Joonho and Zhai, Huanchen and Tong, Yu and Dalzell, Alexander M. and Kumar, Ashutosh and Helms, Phillip and Gray, Johnnie and Cui, Zhi-Hao and Liu, Wenyuan and Kastoryano, Michael and Babbush, Ryan and Preskill, John and Reichman, David R. and Campbell, Earl T. and Valeev, Edward F. and Lin, Lin and Chan, Garnet Kin-Lic},
  journal = {Nature Communications},
  year = {2023},
  volume = {14},
  number = {1},
  pages = {1952},
  doi = {10.1038/s41467-023-37587-6},
}

@article{physxgoogle2025,
    author = {Guang Hao Low and Robbie King and Dominic W. Berry and Qiushi Han and A. Eugene DePrince and Alec F. White and Ryan Babbush and Rolando D. Somma and Nicholas C. Rubin},
    title = {Fast Quantum Simulation of Electronic Structure by Spectral Amplification},
    journal = {Physical Review X},
    year = {2025},
    volume = {15},
    doi = {DOI: https://doi.org/10.1103/pb2g-j9cw}
}

@article{multi-stateiqcc2026,
author = {Lang, Robert A. and Mehendale, Shashank G. and Ryabinkin, Ilya G. and Izmaylov, Artur F.},
title = {Multistate Iterative Qubit Coupled Cluster (MS-iQCC): A Quantum-Inspired, State-Averaged Approach to Ground- And Excited-State Energies},
journal = {Journal of Chemical Theory and Computation},
volume = {22},
number = {4},
pages = {1714-1726},
year = {2026},
doi = {10.1021/acs.jctc.5c01849},
    note ={PMID: 41636614},

URL = { 
    
        https://doi.org/10.1021/acs.jctc.5c01849
    
    

},
eprint = { 
    
        https://doi.org/10.1021/acs.jctc.5c01849
    
    

}

}

@misc{zhai2026classicalsolutionfemocofactormodel,
      title={Classical solution of the FeMo-cofactor model to chemical accuracy and its implications}, 
      author={Huanchen Zhai and Chenghan Li and Xing Zhang and Zhendong Li and Seunghoon Lee and Garnet Kin-Lic Chan},
      year={2026},
      eprint={2601.04621},
      archivePrefix={arXiv},
      primaryClass={physics.chem-ph},
      url={https://arxiv.org/abs/2601.04621}, 
}

@misc{miller2026_trusts,
      title={Approximate simulation of complex quantum circuits using sparse tensors},
      author={Benjamin N. Miller and Peter K. Elgee and Jason R. Pruitt and Kevin C. Cox},
      year={2026},
      eprint={2602.04011},
      archivePrefix={arXiv},
      primaryClass={quant-ph},
      url={https://arxiv.org/abs/2602.04011},
}

@misc{brower2025_blackwell_dmrg,
      title={Mixed-precision ab initio tensor network state methods adapted
             for {NVIDIA} {Blackwell} technology via emulated {FP64} arithmetic},
      author={Cole Brower and Samuel Rodriguez Bernabeu and Jeff Hammond
              and John Gunnels and Sotiris S. Xantheas and Martin Ganahl
              and Andor Menczer and {\"O}rs Legeza},
      year={2025},
      eprint={2510.04795},
      archivePrefix={arXiv},
      primaryClass={physics.chem-ph},
      url={https://arxiv.org/abs/2510.04795},
}

@misc{brown2025_multigpu_network,
      title={Multi-{GPU} Quantum Circuit Simulation and the Impact
             of Network Performance},
      author={W. Michael Brown and Anurag Ramesh and Thomas Lubinski
              and Thien Nguyen and David E. Bernal Neira},
      year={2025},
      eprint={2511.14664},
      archivePrefix={arXiv},
      primaryClass={quant-ph},
      url={https://arxiv.org/abs/2511.14664},
}

@misc{rudolph2025_2dtn,
      title={Simulating and Sampling from Quantum Circuits with {2D}
             Tensor Networks},
      author={Manuel S. Rudolph and Joseph Tindall},
      year={2025},
      eprint={2507.11424},
      archivePrefix={arXiv},
      primaryClass={quant-ph},
      url={https://arxiv.org/abs/2507.11424},
}

@article{carreras2025_vqe_limitations,
      title={Limitations of Quantum Hardware for Molecular Energy
             Estimation Using {VQE}},
      author={Abel Carreras and David Casanova and Rom{\'a}n Or{\'u}s},
      year={2025},
      journal={Phys. Chem. Chem. Phys.},
      doi={10.1039/D5CP03907J},
}

@misc{morchen2024_classification,
      title={Classification of electronic structures and state preparation
             for quantum computation of reaction chemistry},
      author={Maximilian M{\"o}rchen and Guang Hao Low and Thomas Weymuth
              and Hongbin Liu and Matthias Troyer and Markus Reiher},
      year={2024},
      eprint={2409.08910},
      archivePrefix={arXiv},
      primaryClass={quant-ph},
      url={https://arxiv.org/abs/2409.08910},
}

@article{caesura2025_ftqc_speedup,
      title={Faster quantum chemistry simulations on a quantum computer
             with improved tensor factorization and active volume compilation},
      author={Athena Caesura and Cristian L. Cortes and William Pol
              and Sukin Sim and Mark Steudtner and Gian-Luca R. Anselmetti and Matthias Degroote and Nikolaj Moll and Raffaele Santagati and Michael Streif and Christofer S. Tautermann},
      year={2025},
      journal={PRX Quantum},
      doi={10.1103/yngp-5fpm},
}

@article{Larocca_2025,
   title={Barren plateaus in variational quantum computing},
   volume={7},
   ISSN={2522-5820},
   url={http://dx.doi.org/10.1038/s42254-025-00813-9},
   DOI={10.1038/s42254-025-00813-9},
   number={4},
   journal={Nature Reviews Physics},
   publisher={Springer Science and Business Media LLC},
   author={Larocca, Martín and Thanasilp, Supanut and Wang, Samson and Sharma, Kunal and Biamonte, Jacob and Coles, Patrick J. and Cincio, Lukasz and McClean, Jarrod R. and Holmes, Zoë and Cerezo, M.},
   year={2025},
   month=mar, pages={174–189} }
\bibliographystyle{unsrt}

\end{document}